\documentclass[12pt,prd,tightenlines,nofootinbib]{revtex4}
\usepackage{bm}
\usepackage{graphics}
\usepackage{rotating}
\usepackage{epsfig}
\begin{document}
\title{Comment on ``Applications of Two-Body Dirac Equations to the
  Meson Spectrum with Three versus Two Covariant Interactions, SU(3)
  Mixing, and Comparison to a Quasipotential Approach'' 
by~H.~W.~Crater and J.~Schiermeyer}
\author{D. Ebert$^{1}$, R. N. Faustov$^{1,2}$  and V. O. Galkin$^{1,2}$}
\affiliation{
$^1$ Institut f\"ur Physik, Humboldt--Universit\"at zu Berlin,
Newtonstr. 15, D-12489  Berlin, Germany\\
$^2$ Dorodnicyn Computing Centre, Russian Academy of Sciences,
  Vavilov Str. 40, 119991 Moscow, Russia}

\begin{abstract}
We present some considerations in connection with the paper
arXiv:1004.2980 [hep-ph] and give a
response to the criticism of the quasipotential approach in our
relativistic quark model.
\end{abstract}

\maketitle

Recently some comments and remarks concerning the quasipotential
approach in quantum field theory were given  in Ref.\cite{1}. In what
follows we would like to present our understanding of the raised
questions. The basic idea of the quasipotential approach \cite{2} to the
two-particle bound state problem is that the three-dimensional
equation for the wave function $\Psi$ contains the same interaction
kernel $V$ (quasipotential) as the Lippmann-Schwinger equation for the
off-mass-shell scattering amplitude $T$ of  two particles: 
\begin{equation}
  \label{eq:1}
  G_f^{-1}({\bf p})\Psi_E({\bf p})=\int\frac{d^3 q}{(2\pi)^3}\Psi_E({\bf q}),
\end{equation}
\begin{equation}
  \label{eq:2}
  T({\bf p},{\bf q};E)=V({\bf p},{\bf q};E)+\int\frac{d^3 k}{(2\pi)^3}
  V({\bf p},{\bf k};E)G_f({\bf k})T({\bf k},{\bf q};E), 
\end{equation}
where $G_f$  is the Green function of two free particles, ${\bf p}$,
${\bf q}$ and ${\bf k}$ are the relative three-momenta of particles
and $E$ is their total energy. The main property of $G_f$ is
that it implements the exact fulfillment of the relativistic
two-particle unitarity. Thus the quasipotential $V$ can be defined and
constructed in terms of the off-shell amplitude $T$ (e.g. in
perturbation theory). The covariant description can be achieved by
using the conserved four-vector of the total momentum $P$ as the
natural projection axis \cite{3} and consequently the invariant
center-of-mass variables. In this way, the so-called local form of the
quasipotential wave equation can be obtained with the invariant
constraints $(pP)=(qP)=0$ ($p$, $q$ are the relative four-momenta),
which mean the vanishing of relative energies in the centre-of-mass
frame and thus define the off-shell amplitude $T({\bf p},{\bf q},E)$
\cite{2,4}. Such type of equation was widely and successfully used to
calculate the spectra of hydrogenic atoms in QED \cite{4,5,6} to a desired
accuracy without any conceptual problems  within perturbation
theory. Excellent agreement with high precision measurements was
found. 

Inspired by this success, we applied the local quasipotential equation
within the constituent quark model in order to study mesons as
two-particle quark-antiquark bound systems. The quark interaction is
described by QCD, which drastically differs from QED due to the
nonlinear gluon self-interaction.  The latter leads to the phenomenon of asymptotic
freedom and, as believed, to the quark and gluon confinement. Actually,
we know very little about the intrinsic properties of both classical and
quantum solutions of chromodynamics, in particular about their genuine
singularities. Perturbation theory in QCD works only at small
distances (large momenta) though even there it is not self-consistent,
since it deals with free coloured quarks and gluons which presumably
do not exist. Thus, contrary to QED, QCD does not provide
straightforward means for constructing the $q\bar q$ scattering  amplitude
and quasipotential. The appropriate solution of this problem is to
develop a ``QCD motivated'' dynamical model which incorporates  the
expected properties of QCD, such as the linear confinement at large
distances and the Coulomb interaction at small ones. In this way we
obtain the well-known Cornell potential which gives a good description
of spin-averaged heavy quarkonium spectra. In such a situation it would
be inadequate to require the reproduction of all QED-like
contributions to the quasipotential. On the other hand, there could be
some additional singular contributions, e.g. the well-known
Coulomb-like term emerging in the string pattern of 
confinement \cite{bali}. So we adopted the approach of constructing the effective
quasipotential  appropriate for describing the observed properties of
hadrons. To this end we represented the quasipotential as a sum of
the one-gluon exchange part with the running coupling constant and the
confining part consisting of scalar and vector contributions. The
structure of the confining part insures fulfillment of requirements of
the heavy quark effective theory (expansion in $1/m_Q$) and of the
flux tube model leading to the vanishing of the spin-dependent chromomagnetic
interaction. This is achieved due to the Pauli term in the
vector confining part which may originate from the instanton t'Hooft
interaction. The scale dependence of the QCD running coupling
$\alpha_s(\mu)$ is known only for large scales $\mu$, in the
ultraviolet. Its continuation to the infrared region for small scales (light
quarks) is not known and requires complementary admissions and a new
parameter. We have chosen the simplest  behavior of $\alpha_s$ (Badalian, Simonov)
which exhibits the so-called ``freezing'' (finiteness of $\alpha_s$ at $\mu = 0$). It
can be easily sewed with the asymptotically free $\alpha_s$ at the scale of the
$c$ quark mass \cite{7} and well agrees with other analogous
parameterizations (Shirkov and Solovtsev,  lattice simulations). It
should be specially emphasized that compared to the mass spectra the
electroweak decay rates of hadrons much stronger depend on details
of the relativistic quark dynamics.

Based on the above discussion, we 
 now briefly address the ``three weaknesses'', claimed in the
 Conclusion of Ref.~\cite{1}, p.24.
 \begin{enumerate}
 \item  It is worth to point out that we use an ``effective'' quasipotential. The
   most important requirement to its construction  is  to properly
   reproduce and explain the hadron phenomenology, namely mass spectra
   of mesons and baryons, Regge trajectories,  decay and production
   rates, properties of exotic mesons, e.g. tetraquarks, and so on . Thus
   the spin-dependent terms are fixed primarily by this requirement
   and are not related to QED.  Moreover,  there is a very
   non-trivial problem of taking into account the retardation effects
   in the confining part of the quasipotential \cite{8}. 
\item  The quasipotential approach yields excellent
  results \cite{6} when applied to the two-body bound systems within
  QED. The statement that ``QCD .... differs only in the replacement of
  the vector potentials of QED with those of QCD'' is misleading (see
  remarks above). At the moment only phenomenological potential models
  are applicable within  QCD (not mentioning QCD sum rules and the
  lattice).                      
\item Contrary to the statement in Ref.~\cite{1} all relations of the
  heavy quark effective theory in 
the $m_Q\to\infty$ limit are exactly satisfied. In the case of QED the Dirac
equation can be directly obtained from the quasipotential Gross
equation \cite{9} in the infinitely heavy particle limit. 
 \end{enumerate}
Concerning the very last statement (p.25) of Ref.~\cite{1}, mentioning
``... there are, in our view, serious ... short comings insofar as
fundamental field theory connections are concerned'', it would be fair
to say that QCD unfortunately has not reached up to now the level of a
sufficiently  developed theory. Nevertheless, the authors
of \cite{1} admit that ``... the works of Ebert, Faustov, and Galkin
... are quite impressive in terms of spectral and decay agreements''. 

In conclusion, we believe that the QCD-motivated quasipotential
approach is a consistent and effective method to describe successfully
a large variety of hadron physics. By this reason, it is a
convenient and powerful theoretical tool to meet the newly expected
experimental  results and challenges of LHC at least until the future
era, when reliable and detailed QCD calculations are available.

\end{document}